\documentclass[twocolumn,amsmath,amssymb,prl,aps,showkeys]{revtex4}
\usepackage{graphicx}
\usepackage{amsmath,amssymb}
\usepackage{mathrsfs}
\usepackage{color}
\usepackage{afterpage}
\usepackage[version=3]{mhchem}
\usepackage[normal]{subfigure}

\begin{document}
\title{Stability and metastability of clusters in a reactive atmosphere: \\ theoretical evidence for unexpected stoichiometries of Mg$_M$O$_x$}
\author{Saswata Bhattacharya, Sergey V. Levchenko, Luca M. Ghiringhelli, and Matthias Scheffler}
\affiliation{Fritz-Haber-Institut der Max-Planck-Gesellschaft, Faradayweg 4-6, D-14195 Berlin, Germany}
\date{\today}
\begin{abstract}
By applying a genetic algorithm and {\em ab initio} atomistic thermodynamics, we identify the stable and metastable compositions and structures of Mg$_M$O$_x$ clusters at realistic temperatures and oxygen pressures. 
%The stable and metastable systems are identified by {\em ab initio} atomistic thermodynamics. 
We find that small clusters ($M\lesssim 5$) are in thermodynamic equilibrium when $x>M$. The non-stoichiometric clusters exhibit peculiar magnetic behavior, suggesting the possibility of tuning magnetic properties by changing environmental pressure and temperature conditions.
Furthermore, we show that density-functional theory (DFT) with a hybrid exchange-correlation (xc) functional is needed for predicting accurate phase diagrams of metal-oxide clusters. Neither a (sophisticated) force field nor DFT with (semi)local xc functionals are sufficient for even a qualitative prediction.

\end{abstract}
\maketitle
\begin{figure}[t!]
\includegraphics[width=0.8\columnwidth,clip]{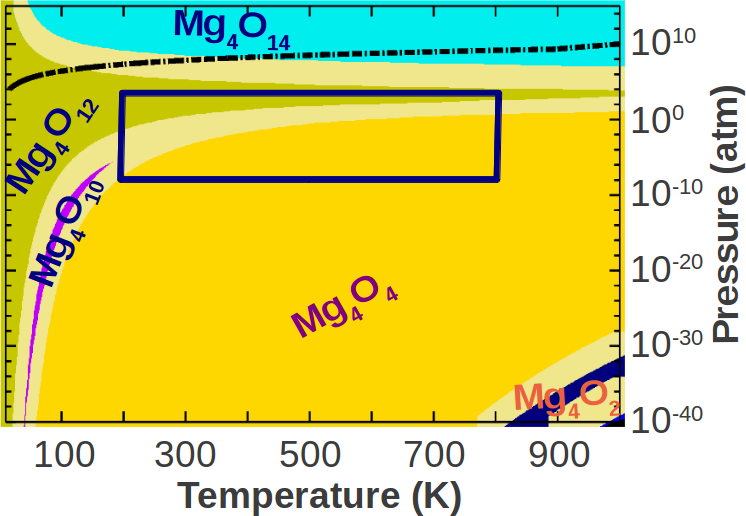}
\caption{Phase diagram for Mg$_4$O$_x$ clusters in an oxygen atmosphere. 
The geometries are optimized with PBE+vdW, (harmonic) free energies are evaluated with the same functional, and the total energies are calculated with PBE0+vdW.
The sand-colored unlabeled areas are regions where different compositions (at least the adjacent ones) coexist (see text). 
The thick black line is the O-rich limit \cite{ms3,aiat}: Above this line, O$_2$ droplets condensate on the clusters.
The rectangle encompasses a region accessible to experiments. }
\label{pt2}
\end{figure}

%%%%%%%%%%%%%%%%%
In the search for novel functional materials, atomic (sub)nanometer clusters are widely studied as model systems exhibiting unique size-dependent properties often even qualitatively different from bulk materials. For example, small clusters may exhibit completely new local structures, stoichiometries, electronic and magnetic properties unknown in the bulk materials~\cite{sasnew1}. Heterogeneous catalysis is just one important example where all mentioned issues are of fundamental importance~\cite{new1, r5, new2, new3, new4, r1, r14n}. \\
\indent The composition and structure of clusters are determined by thermodynamics and kinetics at the relevant temperature ($T$) and the nature of the environment. In thermodynamic equilibrium, only structures and {\em compositions} that minimize the free energy of the combined gas+cluster system will be stable. Although a system is often not in thermodynamic equilibrium, thermodynamic phase diagrams serve as guidelines and important limits for predicting properties and functions of real materials.
In this Letter, we address the issue of stability and metastability using a model system that is relevant for many practical applications: free metal (Mg) clusters in an oxygen atmosphere.\\
\begin{figure}[t]
\includegraphics[width=0.50\textwidth,clip]{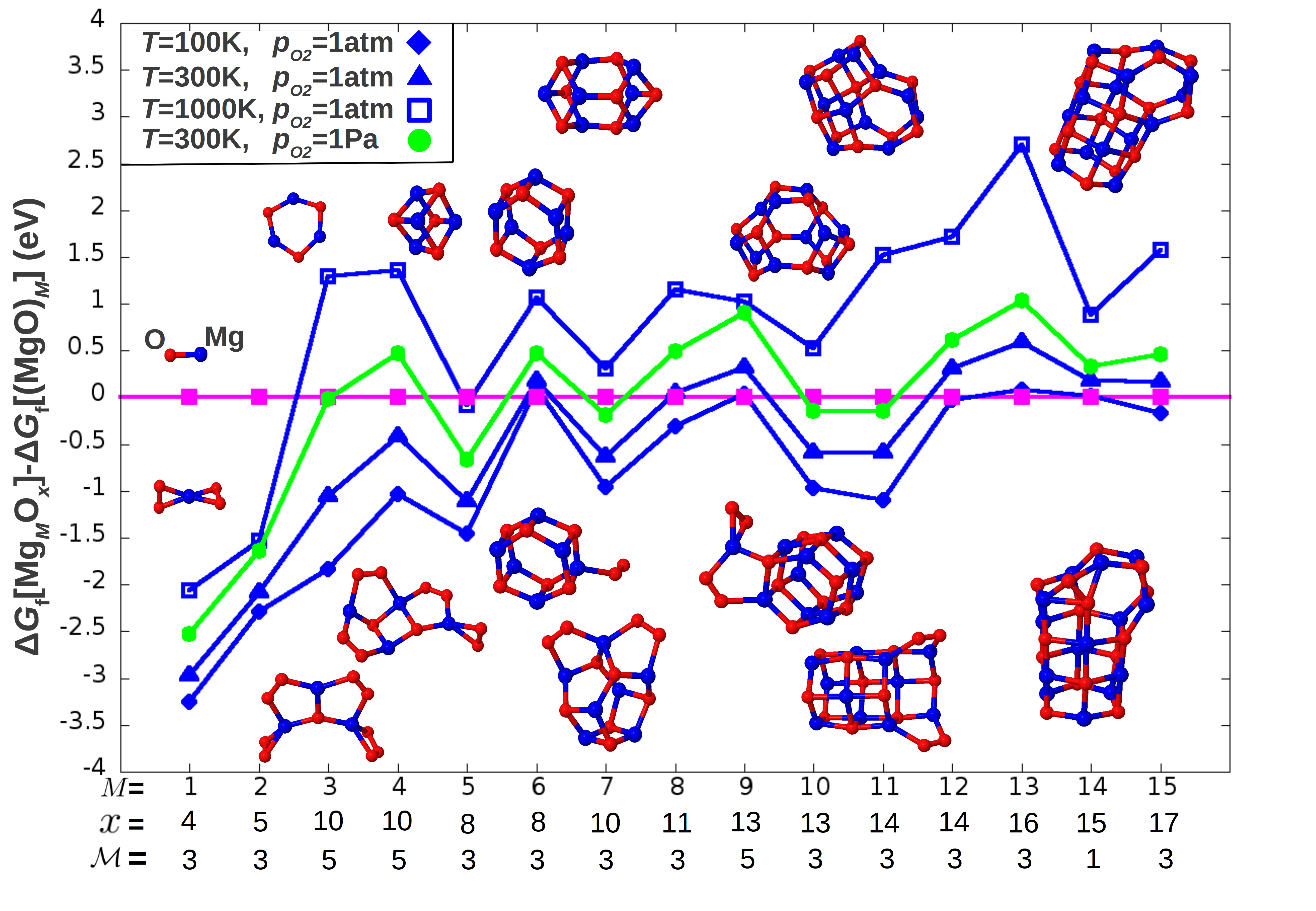}\\
\caption{Free energy of formation of thermodynamically most stable non-stoichiometric clusters (Mg$_M$O$_x$ with $M \neq x$), relative to stoichiometric ($M = x$) clusters, at several $(T,p_{\textrm{O}_2})$ conditions. The geometries were optimized with PBE+vdW, the (harmonic) vibrational free energy was evaluated with the same functional, and the electronic energy was calculated using PBE0+vdW. The label of the horizontal axis shows the amount M of Mg atoms, and the amount x of O atoms and spin multiplicity $\mathcal{M}$ for the non-stoichiometric Mg$_M$O$_x$ cluster thermodynamically most stable at $p_{\textrm{O}_2} = 1$ atm and T = 300 K. (the stoichiometric clusters are all singlets, i.e., $\mathcal{M}=1$). The structure of clusters at selected sizes is also shown (all structures are shown in Suppl. Material)}
\label{comp}
\end{figure}
\indent Most of the previous research on clusters focused on properties of stoichiometric (Mg$_M$O$_x$, $x=M$) clusters~\cite{r1,r4,zpd92,zpd93,r14c,r13,r14e,r7,r10,r6}, and only few attempts have been made to study properties of the non-stoichiometric ($x \neq M$) clusters~\cite{r4,r7,r14n}. However, the decisive issue of stability and metastability of clusters with different compositions at realistic conditions (exchange of atoms with an environment) has not been addressed so far. 
We consider a wide range of Mg$_M$O$_x$ cluster sizes: $1 \! \leq \! M \! \leq \! 15$ and $x$ determined by thermal equilibrium with the environment at given temperature $T$ and partial oxygen pressure $p_{\textrm{O}_2}$. For each stoichiometry, the energy is minimized with respect to both geometry {\em and spin state}. 
Unexpectedly, our results reveal (see Fig. \ref{comp}) that non-stoichiometric clusters with $x > M$ are more stable at realistic ($T, p_{\textrm{O}_2}$ ) when $M < 5$, while for bigger clusters there is a competition between stoichiometric and more-than-stoichiometric composition ($x > M$).\\ 
\indent For the determination of low-energy structures, we use a genetic algorithm (GA). GA mimics the process of ``natural'' selection to evolve a pool of atomic structures until the structures that fit best chosen selection criteria are found~\cite{r15,r16,r17}. In this work, we search for structures that minimize the DFT total energy within each stoichiometry. Note that GA is not a single method, but a whole class of methods, and must be optimized and {\em validated} for each system. Details of our implementation of GA are given in the Suppl. Material, and its validation is discussed below.\\
\indent The free energy as a function of $T$ and $p_{\textrm{O}_2}$ is calculated for the minimum of the potential-energy surface (global minimum, GM) and (energetically) adjacent local energy minima for each stoichiometry using the {\em ab initio} atomistic thermodynamics ({\em ai}AT) approach~\cite{ms1,ms2,ms3,aiat}, recently extended to cluster systems~\cite{lmg1,eli}. The thermodynamic phase diagram is then constructed by selecting cluster compositions and structures with the lowest free energy as a function of ($T$, $p_{\textrm{O}_2}$). \\
\begin{figure} [t!]
\includegraphics[width=0.85\columnwidth,clip]{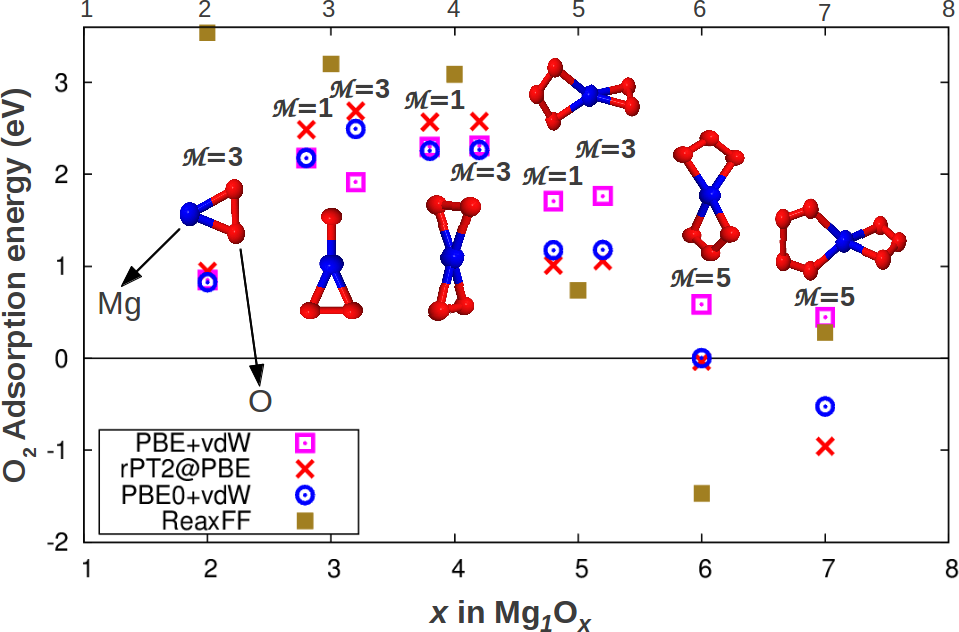}
\caption{Energy of O$_2$-adsorption on MgO$_x$ clusters (energy of the reaction MgO$_x$ + O$_2$ $\rightarrow$ MgO$_{x+2}$ calculated at the PBE+vdW GM geometry) calculated with different xc functionals and reaxFF~\cite{ff}. For the cases where singlet ($\mathcal{M}=1$) and triplet ($\mathcal{M}=3$) spin states are almost degenerate, the two sets of energies are shown (for DFT only, since reaxFF does not describe spin). In these cases, the spin state of MgO$_x$ is chosen to be singlet for triplet MgO$_{x+2}$, and triplet for singlet MgO$_{x+2}$, to preserve the spin-conservation rule (O$_2$ is always kept in its ground triplet electronic state). The geometry differences of clusters with different spin states are invisible at this scale.}
\label{be2}
\end{figure}
\indent Reliability of predictions on relative stability of clusters with different structures depends on the accuracy of the underlying potential-energy surface (PES). Here, we find that comparing clusters with different stoichiometry poses an additional challenge, which is not apparent if only one stoichiometry is considered. Fig.~\ref{be2} shows a comparison of DFT energies of the reaction MgO$_x$ + O$_2$ $\rightarrow$ MgO$_{x+2}$ calculated with different xc functionals: generalized gradient approximation PBE \cite{pbe} and hybrid PBE0 \cite{pbe0}, both corrected for the van-der-Waals (vdW) interaction (PBE+vdW and PBE0+vdW), and the highest level currently achievable within the DFT framework, i.e., the renormalized second-order perturbation theory (rPT2)~\cite{xinguo}, here applied on PBE orbitals (rPT2@PBE). All DFT calculations are performed with the all-electron FHI-aims package, employing numerically-tabulated atomic orbitals~\cite{aims}. It can be seen that PBE+vdW strongly overestimates stability of clusters with larger $x$, resulting in a qualitatively incorrect prediction that O$_2$ adsorption would be favored over desorption up to a large excess of oxygen. Such behavior is not confirmed by the PBE0+vdW hybrid functional or rPT2@PBE (note that a similar behavior is found for PBE compared to PBE0, i.e., without vdW correction).\\
\indent Interestingly, for lower O$_2$-coverage the difference between PBE+vdW and PBE0+vdW/rPT2@PBE energies of O$_2$ adsorption on Mg and MgO$_2$ is small despite the error in the O$_2$ binding energy (see Fig.~\ref{be2}; the calculated O$_2$ binding energy is 6.23 eV for PBE+vdW, 5.36 for PBE0+vdW, 4.59 for rPT2@PBE, and the experimental value is 5.21~\cite{exp-o2}). This can be explained by error cancellation for the clusters: If an O$_2$ molecule adsorbs non-dissociatively on Mg$_M$O$_x$, the error in the description of Mg$_M$O$_{x+2}$ will cancel with the error in the description of O$_2$ when calculating the adsorption energy. Indeed, we find that adsorption of O$_2$ on Mg and MgO$_2$ does not lead to breaking of O-O bonds. The difference between PBE+vdW and PBE0+vdW energies of O$_2$ adsorption on MgO for the triplet case is due to the difference in the description of the singlet state of MgO itself. For clusters with $x\geq 5$, correction of the O$_2$ binding energy error with the experimental value increases the difference between PBE+vdW and PBE0+vdW/rPT2@PBE adsorption energies (see the Suppl. Material). The tendency of PBE (and, as a consequence, of PBE+vdW) to overbind O$_2$ molecules at high coverage holds also at larger $M$ (see the Suppl. Material).\\
\indent It has been recently shown~\cite{norina} that HSE06 functional~\cite{hse06} yields a good description of the level alignment and electron transfer in MgO. Based on comparison for selected clusters, we find that PBE0, which belongs to the same family of functionals, yields formation energies essentially identical to HSE06 (within 0.05 eV for Mg$_M$O$_x$ neutral clusters). Taking into account this comparison and the fact that PBE0+vdW results are in general much closer to rPT2@PBE than PBE+vdW, we conclude that reducing the self-interaction error by including Hartree-Fock exchange in DFT is crucial for correct description of the Mg$_M$O$_x$ cluster energetics, both within each stoichiometry and in particular when comparing different stoichiometries.\\
\indent Therefore, the PBE0+vdW functional is used to calculate the energies and evaluate the fitness during GA runs. In order to improve the efficiency of the GA scan, our implementation proceeds in terms of a cascade in which successive steps employ higher levels of theory, with each next level using information obtained at the lower level.  Initially, a local optimization of a given structure is performed with lower-level (computationally relatively cheap) numerical settings. At this level, PBE+vdW and ``light" numerical settings with basis set ``tier 1'' were used~\cite{aims}, and forces were converged to better than 10$^{-3}$ eV/\AA. Next, structures with energies within 2.5~eV from the current GM candidate, are further relaxed using higher-level settings~\cite{note:otherga}. We use PBE+vdW functional with ``tight - tier 2'' numerical and basis settings for energy minimization at the higher level, and forces were converged to better than 10$^{-5}$ eV/\AA. Next, the energy of these further optimized structures are re-evaluated using PBE0+vdW and ``tight - tier 2'' settings.\\%~\cite{note:hlev}.\\
\indent A challenging problem of any random-walk-type multi-dimensional global minimization scheme (including basin hopping and GA) is to guarantee that the lowest-energy structure found by the algorithm is indeed the GM. We address this validation problem by applying replica-exchange molecular dynamics (REMD)~\cite{remd1,remd4,lmg1}, a (computationally very expensive) reference method that performs the canonical sampling of a PES simultaneously at different temperatures, and is exhaustive if the system is ergodic. The validation of GA is performed using a reactive force-field (reaxFF \cite{ff,ffMgO}) to evaluate energy and forces. While the force-field is by far not accurate enough to predict correct energy differences, as can be seen in Fig.~\ref{be2} and in Suppl. Material, using it for the validation is a more stringent test for the GA, since the force-field PES is found to have a much more complicated landscape, with many more local minima, than the {\em ab initio} PES. 
With our implementation of GA we find, for the systems here considered, the same GM as found by 1.5-$\mu$s-long (cumulative time) REMD~\cite{long} runs. An independent evidence of the robustness of our GA scheme is that, for stoichiometric clusters, we could always find the GM reported in literature \cite{r5,r6}.\\
\indent At given $T$, $p_{\rm O_2}$, and $M$, the stable stoichiometry of a Mg$_M$O$_x$ cluster is determined via {\em ai}AT, i.e., by minimizing the Gibbs free energy of formation \cite{eli}: $ \Delta G_f(T, p_{\textrm{O}_2}) = F_{\textrm{Mg}_M\textrm{O}_x} (T) - F_{\textrm{Mg}_M} (T) - x\mu_\textrm{O} (T, p_{\textrm{O}_2}) $. Here, $ F_{\textrm{Mg}_M\textrm{O}_x} (T) $ and $F_{\textrm{Mg}_M} (T)$ are the Helmholtz free energies of the Mg$_M$O$_x$ and the pristine Mg$_M$ cluster (at their ground state with respect to geometry and spin), respectively, and $ \mu_\textrm{O} (T, p_{\textrm{O}_2})$  is the chemical potential of oxygen.
$ F_{\textrm{Mg}_M\textrm{O}_x} (T) $ and $F_{\textrm{Mg}_M} (T)$ are calculated using DFT information and are expressed as the sum of DFT total energy, DFT vibrational free energy in the harmonic approximation, as well as translational, rotational, symmetry- and spin-degeneracy free-energy contributions. The dependence of $ \mu_\textrm{O}$ on $T$ and $p_{\rm O_2}$ is calculated using the ideal (diatomic) gas approximation with the same DFT functional as for the clusters~\cite{eli}. Note that $\mu_\textrm{O} (T, p_{\textrm{O}_2})$ is in turn a sum \cite{eli} of the same free-energy contributions as for the free clusters, where the $p_{\textrm{O}_2}$ dependence is captured by the translational free energy term, i.e., $k_{\rm B} T \ln p_{\textrm{O}_2} + f(T)$ (where the second term does not depend on $p_{\textrm{O}_2}$). 
The phase diagram for a particular $M$ is constructed by identifying the lowest-free-energy structures at each ($T$, $p_{\textrm{O}_2}$). As a representative example, we show in Fig.~\ref{pt2} the phase diagram for $M$ = 4. 
Phase diagrams based on reaxFF, PBE+vdW, PBE0 (without vdW), and rPT2@PBE can be found in the Suppl. Material. We find that at all DFT levels, the phase diagrams are qualitatively and quantitatively very similar for $T > 200$ K and $p_{\textrm{O}_2} < 10^{-5}$ atm. For higher pressures and/or lower temperatures, however, PBE+vdW predicts larger $x$ in Mg$_M$O$_x$ as thermodynamically more stable, compared to PBE0+vdW and rPT2@PBE. This is consistent with the results shown in Fig. \ref{be2}, i.e., PBE tends to favor adsorption of a larger number of O$_2$ molecules.
PBE0 and PBE0+vdW yield almost identical phase diagrams. Thus, at the considered cluster sizes the vdW interactions, within the scheme of Ref. \cite{ts-scheme}, do not affect the differences between free energies of formation of competing Mg$_M$O$_x$ clusters. ReaxFF-based diagrams were evaluated for the sake of comparison: they are very different from DFT-based ones and no conclusions, even qualitative, can be drawn from the analysis of the reactive force field results. \\
\indent For creating the phase diagrams, we have approximated the configurational free energy with the harmonic vibrational free energy. If the clusters exhibit fluxional behavior or are melted, as it is the case for some metal clusters, this approximation can become invalid at high or even moderate temperatures~\cite{lmg1,confent}. For all the Mg$_M$O$_x$ clusters described here, we have tested the validity of the harmonic assumption by running {\em ab initio} MD simulations for 20 ps at $T=800$ K \cite{note:MD}. We found that, while the structures exhibit large vibrations, their topology is never destroyed, and that the initial 0 K structure was reversibly recovered by cooling down the hot sample. At most, some of the $O_2$ moieties (see below) either freely rotate or roam around the clusters at larger $x$. In order to quantitatively account for these larger displacements in term of configurational entropy, we have evaluated the excess free energy of selected clusters, by integrating the thermodynamic relation $\partial (\beta F(\beta)) / \partial \beta = \langle U \rangle_{\beta} $, where $\langle U \rangle_{\beta}$ is the canonical average of the total energy at temperature $T = 1 / k_{\rm B} \beta$. The differential equation is numerically integrated by evaluating $\langle U \rangle_{\beta}$ via 4-ps-long (after equilibration) MD trajectories at five increasing temperatures, from $T=10$~K (where the reference free energy is safely approximated by the harmonic expression) to $T=800$~K. The integration path was checked to be reversible. In this way all anharmonic contributions are included through the canonical sampling of the PES. 
For the Mg$_4$O$_x$, the anharmonic correction enlarges the stability region at higher temperatures of the highest stoichiometry, Mg$_4$O$_{12}$, by at most 50~K at $p_{\textrm{O}_2} = 1$~atm.
Note that the (harmonic or beyond) configurational free energy is only one part of the total free energy and that, at higher $(T$, $p_{\textrm{O}_2})$ and stoichiometries $x$, the free energy is dominated by the term $x k_{\rm B} T \ln p$. For this reason, the plots are extended up to 1\,000 K, where weakly bound O$_2$ moieties may evaporate; nonetheless, at thermodynamic equilibrium the pressure of O$_2$ molecules ensures that another O$_2$ molecule will on average replace the evaporated one. Furthermore, we find (see Suppl. Material) that neglecting translational, rotational, (harmonic or beyond) vibrational, and symmetry- and spin degeneracy-related free-energy contributions results only in slight changes in the phase diagrams, comparable to the differences between PBE0+vdW and rPT2@PBE diagrams, for all considered $M$.\\
\indent We have constructed phase diagrams for $1 \! \leq M \! \leq \! 15 $. At thermodynamic equilibrium in a wide range around normal conditions, small Mg$_M$O$_x$ clusters ($M \leq 5)$ are found to be preferentially non-stoichiometric ($x > M$). For sizes $M > 5$ we observe a competition between stoichiometric and non-stoichiometric stable structures.
In Fig.~\ref{comp}, we also report the relative stability of stoichiometric and non-stoichiometric structures at low ($T= 100$~K) and high ($T=1\,000$~K) temperatures. The relative stability of stoichiometric structures increases with temperature. 
\indent A further interesting property we find is that Mg$_M$O$_x$ non-stoichiometric clusters have in general several near-degenerate (within $\sim$0.05~eV) electronic states with different spin. 
The highest multiplicity we predict is $\mathcal{M} = 7$ for the case Mg$_4$O$_{12}$, and values of $\mathcal{M}=3$ and 5 are largely represented. 
The analysis of the spin density shows that the unpaired electron density is localized on O$_2$ and O$_3$ moieties (see examples in the Suppl. Material).
The orbitals occupied by the different unpaired electrons in the same cluster have vanishing overlaps due to large separation.
The different spin-states for each isomer can be formed and coexist in oxygen atmosphere without violating spin conservation rules, since successive adsorption and desorption of the (triplet) O$_2$ molecules in the gas phase allows the clusters to reach all the stable spin-multiplicities observed for each cluster. The thermodynamically favored access of oxygen in small clusters is in sharp contrast to bulk pristine MgO, where stoichiometric composition is strongly favored. In all cases where O$_2$ or O$_3$ moieties are coordinated to maximum two Mg atoms, these moieties host an unpaired spin. When the number of Mg is increased such that every O atom can be coordinated to three or more Mg, and at the same time the clusters have large HOMO-LUMO gap (i.e., all valence Mg electron are transferred to O$_x$), then stoichiometric clusters become thermodynamically stabilized. \\
\indent All the quasi-degenerate spin states are populated at finite temperature, thus all the non-stoichiometric clusters with energetically quasi-degenerate states are paramagnetic. In contrast to non-stoichiometric clusters, we find that stoichiometric structures are always singlet, separated from the higher-multiplicity states by at least 1~eV. Thus, the stoichiometric clusters are diamagnetic.
By looking at Fig.~\ref{pt2}, we note that in the range of realistically achievable pressures encompassed by the rectangle, non-stoichiometric (paramagnetic) structures are thermodynamically more stable at lower temperatures, while at higher temperatures the stoichiometric (diamagnetic) structures become more stable. The temperature at which this transition occurs is a rapidly varying function of $p_{\textrm{O}_2}$. We have thus identified a class of systems that undergo an unusual paramagnetic-diamagnetic transition induced by $T$ and $p$ of the reactive atmosphere, where the change in magnetic behavior reflects the change in composition. 
The transition between paramagnetic and diamagnetic behavior is smooth when environmental conditions are changed smoothly, because different stoichiometries always coexist within few $k_{\rm B}T$. In fact, the sand-colored areas in Fig.~\ref{pt2} are the regions where the free energies of the competing compositions/structures are within an energy range of $2k_\textrm{B}T$.\\
\indent In conclusion, we have presented a theoretical framework for predicting structure and stoichiometry of stable and metastable clusters in thermodynamic equilibrium with a gas atmosphere. An efficient and unbiased scan of the potential energy of the clusters at various compositions is combined with {\em ab initio} atomistic thermodynamics, a tool for evaluating the relative free energy of structures by knowing their electronic energy, vibrational frequencies, and structural parameters for the evaluation of translational and rotational entropy.
The methodology has been applied to Mg clusters in an oxygen atmosphere. We have shown that small alkaline earth metal clusters form thermodynamically stable non-stoichiometric ``nano-oxides", which have been overlooked so far.  They are also predicted to have peculiar $(T,p)$-dependent, magnetic properties.\\
\indent We acknowledge the cluster of excellence ``Unifying Concepts in Catalysis" (UniCat, sponsored by the DFG and administered by the TU Berlin) for financial support.


\begin{thebibliography}{99}
\bibitem{sasnew1} E. Roduner, Chem. Soc. Rev. {\bf 35}, 583 (2006). 
\bibitem{new1} K. G. Caulton, M. G. Thomas, B. A. Sosinsky, and E. L. Muetterties, Proc. Natl. Acad. Sci. USA {\bf 73} (1976).
\bibitem{r5} K. Kwapien, M. Sierka, J. Dobler, J. Sauer, M. Haertelt, A. Fielicke, and G. Meijer, Angew. Chem. Int. Ed. {\bf 50}, 1716 (2011).
\bibitem{new2} N. Marom, M. Kim, and J. R. Chelikowsky, Phys. Rev. Lett. {\bf 108}, 106801 (2012).
\bibitem{new3} Y. Gao, N. Shao, Y. Pei, Z. Chen, and X. C. Zeng, ACS Nano {\bf 5}, 7818 (2011).
\bibitem{new4} Yimin Li \emph{et. al.}, J. Am. Chem. Soc. {\bf 133}, 13527 (2011).
\bibitem{r1} W. A. Saunders, Phys. Rev. B {\bf37}, 6583 (1988).
\bibitem{r14n} T. Uchino, T. Yoko Phys. Rev. B {\bf85}, 012407 (2012).
\bibitem{r4} P. J. Ziemann and A. W. Castleman, Jr., Phys. Rev. B {\bf44}, 6488 (1991).
\bibitem{zpd92} S. Moukouri, C. Noguera Z.Phys. D {\bf 24}, 71 (1992)
\bibitem{zpd93} S. Moukouri, C. Noguera Z.Phys. D {\bf 27}, 79 (1993)
\bibitem{r14c} J.M. Recio, R. Pandey, Phys. Rev. A {\bf47}, 2075 (1993).
\bibitem{r13}  M.-J. Malliavin, C. Coudray, J. Chem. Phys. {\bf106}, 2323 (1997).
\bibitem{r14e} E. de la Puente, A. Aguado, A. Ayuela, J.M. Lopez, Phys. Rev. B {\bf56}, 7607 (1997).
\bibitem{r7} C. Roberts, R.L. Johnston, Phys. Chem. Chem. Phys. {\bf3}, 5024 (2001).
\bibitem{r10} A. Jain, V. Kumar, M. Sluiter, and Y. Kawazoe, Comp. Mater. Sc. {\bf36}, 171 (2006).
\bibitem{r6} M. Haertelt, A. Fielicke, G. Meijer, K. Kwapien, M. Sierkaz and J. Sauer, Phys. Chem. Chem. Phys., {\bf14}, 2849 (2012).
\bibitem{r15} R. L. Johnston, Dalton Trans. {\bf 22}, 4193 (2003).
\bibitem{r16} M. Sierka, Progress in Surface Science {\bf85}, 398 (2010).
\bibitem{r17} L. B. Vilhelmsen and B. Hammer, Phys. Rev. Lett. {\bf108}, 126101 (2012).
\bibitem{long} S. Bhattacharya, S. V. Levchenko, L. M. Ghiringhelli, M. Scheffler, to be submitted (2013).
\bibitem{crossover} D. M. Deaven and K. M. Ho, Phys. Rev. Lett., {\bf75}, 288 (1995).
\bibitem{ms1} C. M. Weinert and M. Scheffler, in Defects in Semiconductors, edited by H. J. von Bardeleben (Mat. Sci. Forum 10–12, 1986), pp. 25–30.
\bibitem{ms2} M. Scheffler and J. Dabrowski, Phil. Mag. A {\bf58}, 107 (1988).
\bibitem{ms3} K. Reuter and M. Scheffler, Phys. Rev. B {\bf65}, 035406 (2001). Erratum: Phys. Rev. B {\bf 75}, 049901(E) (2007). 
\bibitem{aiat} K. Reuter, C. Stampfl, and M. Scheffler, in Handbook of Materials Modeling (2005 Springer), edited by S. Yip, pp. 149–194.
\bibitem{lmg1} E. C. Beret, L. M. Ghiringhelli, and M. Scheffler, Faraday Discuss. {\bf152}, 153 (2011).
\bibitem{eli} E. C. Beret, M. van Wijk, and L. M. Ghiringhelli. To be published on Int. J. Quantum Chem. (2013). (published online, DOI: 10.1002/qua.24503 )
\bibitem{pbe} J. P. Perdew, K. Burke, and M. Ernzerhof, Phys. Rev. Lett. {\bf77}, 3865 (1996). Erratum: J. P. Perdew, K. Burke, and M. Ernzerhof, Phys. Rev. Lett. {\bf78}, 1396 (1997).
\bibitem{pbe0} J. P. Perdew, K. Burke, and M. Ernzerhof, Phys. Rev. Lett. {\bf 77}, 3865 (1996). 
%\bibitem{note:hlev} %The difference in binding energy between an isomer optimized with PBE0+vdW forces (tight / tier 1 / forces converged to 10$^{-5}$ eV/\AA), and the same isomer optimized with PBE+vdW (tight / tier 2 / forces converged to 10$^{-5}$ eV/\AA), when the energy of both geometries is evaluated via PBE0+vdW (tight / tier 2), is small, i.e., at most 0.04~eV among all cases we checked. Thus, the computational cost of the PBE0+vdW further optimization is not justified (we estimated a gain of up to a factor 2 of overall computational time just by skipping the latter optimization).
\bibitem{xinguo} X. Ren, A. Tkatchenko and M. Scheffler, Phys. Rev. Lett. {\bf 106}, 153003 (2011). X. Ren, P. Rinke, C. Joas, and M. Scheffler, J. Mater. Sci. {\bf 47}, 7447 (2012).
%\bibitem{note:rpt2lev} rPT2@PBE energies were calculated with FHI-aims, using ``really tight" grid settings and ``tier 4" basis set, and were counterpoise-corrected for the basis set superposition error.
\bibitem{aims} V. Blum, R. Gehrke, F. Hanke, P. Havu, V. Havu, X. Ren, K. Reuter, and M. Scheffler, Comp. Phys. Comm. {\bf180}, 2175 (2009).
PBE and PBE0 energies were calculated using ``tight - tier 2'' numerical and basis settings. rPT2@PBE energies were calculated using ``really tight" grid settings and ``tier 4" basis set, and counterpoise-corrected for the basis set superposition error.
\bibitem{exp-o2} Chase Jr., M. W., NIST-JANAF Tables ``4th ed.'', J. Phys. Chem. Ref. Data Monogr. 9, Suppl. 1 (1998). Zero-point energy correction taken from Huber, K. P. and Herzberg, G., in {\em Molecular Spectra and Molecular Structure: Constants of Diatomic Molecules} (Van Nostrand Reinhold, New
York, 1979), Vol. 4.
\bibitem{norina} N. Richter, S. Sicolo, S. Levchenko, J. Sauer, and M. Scheffler. To be published on Phys. Rev. Lett. (2013). 
\bibitem{hse06} A. V. Krukau, O. A. Vydrov, A. F. Izmaylov, and G. E. Scuseria, J. Chem. Phys.  {\bf 125}, 224106 (2006).
\bibitem{note:otherga} A two-level GA was proposed in Ref. \cite{r17}, however our cascade approach is developed into a wider scheme (see Suppl. Material).
\bibitem{remd1} E. Marinari, G. Parisi, Europhys. Lett. {\bf 19}, 451 (1992). 
\bibitem{remd4} D. J. Sindhikara, D. J. Emerson, and A. E. Roitberg, J. Chem. Theory Comput. {\bf 6}, 2804 (2010). 
\bibitem{ff} A.C.T. van Duin, S. Dasgupta, F. Lorant and W. A. Goddard,  J. Phys. Chem. A {\bf 105}, 9396 (2001).
\\ The reaxFF parameters for Mg and O \cite{ffMgO} were fit \cite{ffMgO} to a quantum-mechanics training set including MgO$-$bulk and O$_2-$molecule properties. We checked that this parametrization yields for bulk MgO lattice constant within 1 \% and bulk modulus within 10 \% from the experimental values; similarly, for the MgO and O$_2$ dimer, it gives bond length and vibrational frequency within 1 \% and 10\% from the respective reference values. On this set of properties, reaxFF performs remarkably well, comparable to PBE+vdW \cite{pbe}, PBE0+vdW \cite{pbe0}, and HSE06+vdW \cite{hse06}, but it fails very clearly for small clusters. In particular, ReaxFF yields a qualitatively good pre-scanning for stoichiometric clusters, while for non-stoichiometric ones it is far from desirable accuracy. This is not unexpected, as not even PBE is able to account for the charge redistributions associated with non-stoichiometry clusters. 
\bibitem{ffMgO} R. Zhu, F. Janetzko, Y. Zhang, A. C. T. van Duin, W. A. Goddard and D. R. Salahub, Theor. Chem. Account, {\bf 120} 479 (2008).
\bibitem{ts-scheme} A. Tkatchenko and M. Scheffler, Phys. Rev. Lett. {\bf102}, 073005 (2009).
\bibitem{confent} P. Koskinen, H. H\"{a}kkinen, B. Huber, B. von Issendorff and M. Moseler, Phys. Rev. Lett. {\bf 98}, 015701 (2007).
\bibitem{note:MD} The Born-Oppenheimer MD simulations were performed with the PBE functional, tight/tier-2 settings, 1 fs time-step and stochastic velocity-rescaling thermostat. 
\end{thebibliography}
\end{document}

%\section*{Supplemental Material}
\begin{center}
{\Large \bf Supplemental Material}\\ 
\end{center}
\vspace*{12pt}
\begin{enumerate}[\bf I.]

\item {\bf Thermodynamic stability of stoichiometric vs non-storichiometric Mg$_M$O$_x$ clusters}

\item {\bf Details on the implemented GA schemes}

\item {\bf Performance of reaxFF}

\item {\bf O$_2$-adsorption energy on MgO$_x$, with functionals corrected by the experimental value of O$_2$ binding energy}

\item {\bf O$_2$-adsorption energy on Mg$_2$O$_x$ and Mg$_3$O$_x$ clusters}

\item {\bf Mg$_2$O$_x$ phase diagrams with various functionals}

\item {\bf Effect of translational, rotational, vibrational contributions to the free energy on Mg$_2$O$_x$ phase diagram}

\item {\bf Effect of anharmonic contributions to configurational free energy}

\item {\bf Examples of spin densities on non-stochiometric Mg$_M$O$_x$}

\end{enumerate}
%\newpage

% \begin{figure*}[t!]
% {\Large \bf Supplemental Material}\\
% \end{figure*}

\begin{figure*}[b!]
{\bf \Large I. Thermodynamic stability of stoichiometric vs non-storichiometric Mg$_M$O$_x$ clusters \\}
%\vspace{24pt}
\includegraphics[width=0.8\columnwidth,clip]{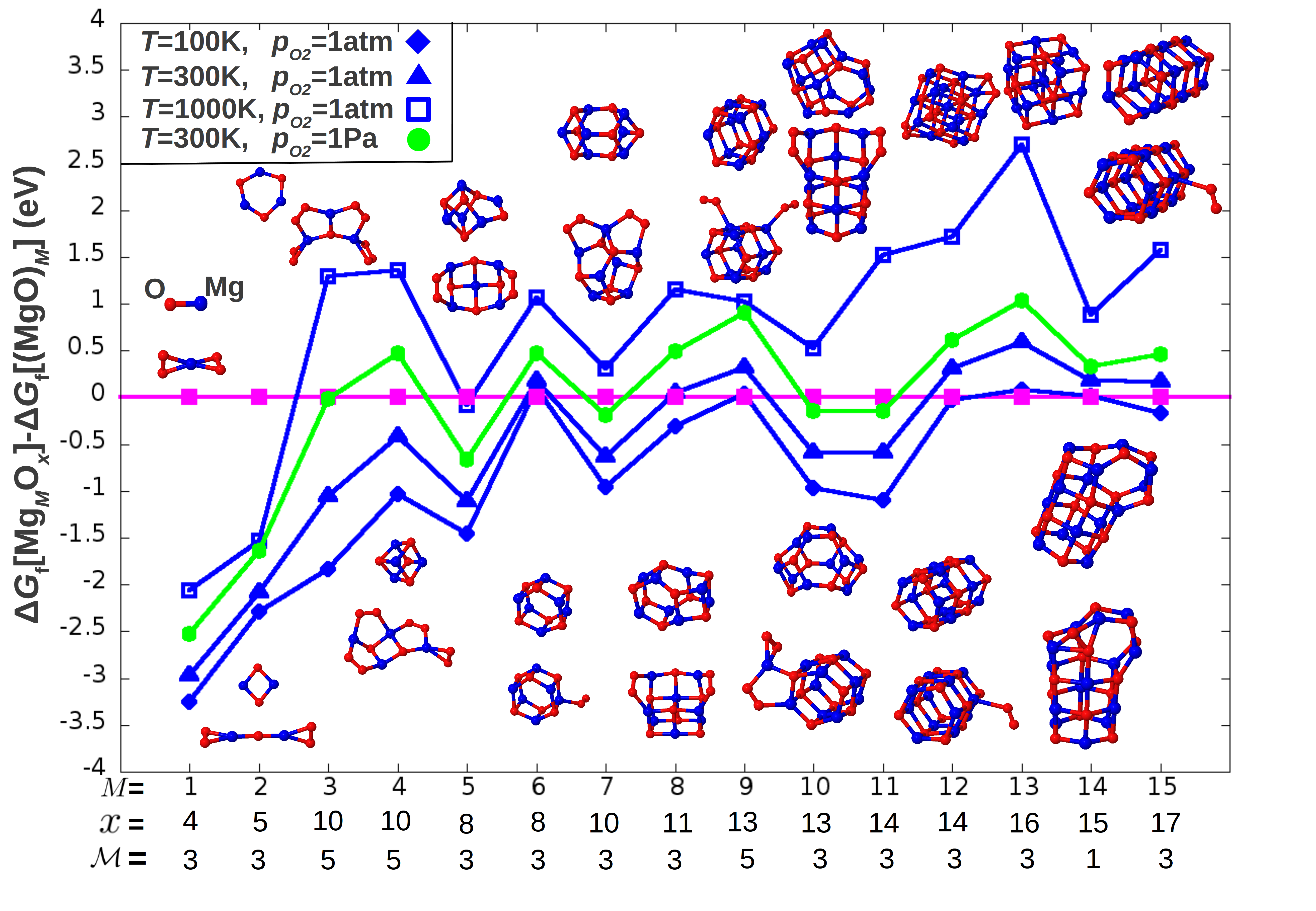}
\caption[]{Free energy of formation of thermodynamically most stable non-stoichiometric (Mg$_M$O$_x$ with $M \neq x$) relative to stoichiometric ($M = x$) clusters at several $(T,p_{\textrm{O}_2})$ conditions. The geometries were optimized with PBE+vdW, and the electronic energy was calculated using PBE0+vdW. The label of the horizontal axis shoes, below the amount $M$ of Mg atoms, the amount $x$ of O atoms for the thermodynamically most stable non-stoichiometric cluster at the corresponding $M$, at $p_{\textrm{O}_2} = 1$~atm and $T=300$ K. For the same thermodynamic condition, the third line reports the spin multiplicity $\mathcal{M}$ of the lowest free-energy non-stoichiometric Mg$_M$O$_x$ (the stoichiometric clusters are all singlets, i.e., $\mathcal{M}=1$).}
\label{SM:I}
\end{figure*} 

\clearpage
\newpage
\begin{center}
{\bf \Large II. Details on the implemented GA scheme \\} {\bf The benchmark and full detail on validation is found in \cite{long}\\} 
\end{center}

Schematically, our cGA algorithm proceeds as follows (all terms in italic will be explained afterwards):
\begin{enumerate}[(1)]
  \item Selection of a composition of the clusters and formation of an initial pool of random structures, locally optimized by a classical force field (FF).
 \item Evaluation of the {\em fitness} function for all structures (using FF binding energy). % \textcolor{red}{Different schemes}
 \item GA global optimization using the classical FF. This consists in the iteration of steps (i)--(v): 
 \begin{enumerate}[(i)]
  \item {\em Selection} of two structures (in GA jargon, {\em parents}).
  \item Assemblage of a trial structure ({\em child}) through {\em crossover} and {\em mutation}.
  \item Local optimization (force minimization) of child structure using the classical FF.
  \item Evaluation of the {\em fitness} function. Comparison of the optimized child with existent structures; reject if {\em similar}, jump to (i). 
  \item Check whether convergence has been reached. If so, stop FF-GA and go the next step, DFT-GA.
 \end{enumerate}
 \item Formation of a new pool of structures using best fit structures from FF-GA, locally optimized at the DFT level (PBE+vdW, {\em low-level settings}).
 \item Calculation of fitness function for all structures (using energy at the PBE0+vdW level).
 \item GA scheme using DFT. In practice iteration of steps (a)--(i):
 \begin{enumerate}[(a)]
  \item {\em Selection} of two structures.
  \item Assemblage of a child structure through {\em crossover} and {\em mutation}.
  \item Local optimization of the child structure with PBE+vdW, {\em low-level settings}.
  \item Comparison of the optimized child with existent structures. {\em Early rejection} if {\em similar}; jump to (a). 
  \item Further local optimization of the child with PBE+vdW, {\em high-level settings}.
  \item Harmonic analysis of the optimized child; if unstable, perturb along the unstable mode and go back to (c). 
  \item Evaluation of {\em fitness} function based on PBE0+vdW total energy. 
  \item Check whether convergence has been reached. If so, stop.
 \end{enumerate}
\end{enumerate}

In the following, we analyze one by one the key words introduced in the detailed scheme above.

% \subsubsection*{\bf Initial random pool}
% \label{rp}
%  We generate structures with atoms randomly distributed on the surface of an ellipsoid with axes of random length, with the constraint that the closest distance between neighbors is larger than a certain threshold (for Mg-O, O-O, and Mg-Mg, the threshold was set to 1.5 \AA). In particular we check that we get structures in three flavors: nearly spherical (three axes of almost equal length), prolate (two short axes, one long) and oblate (two long axes and one short). Note that the limit of prolate structures are linear structures and the limit of oblate ones are planar structures.\\ 
 
\subsubsection*{\bf Fitness function}
\label{ff}
Each cluster $i$ in the population is assigned a normalized fitness value, $\rho_i$, based on its total energy (binding energy for the FF):
\begin{equation}
\label{eqn1}
\rho_i=\frac{\epsilon_i}{\sum_i\epsilon_i}
\end{equation}
and $\epsilon_i$ is the relative energy of the $i^{th}$ cluster as defined below:
\begin{equation}
\label{eqn2}
\epsilon_i=\frac{E_\textrm{max}-E_i}{E_\textrm{max}-E_\textrm{min}}
\end{equation}
Where $E_i$ is the total energy of the $i^{th}$ cluster of the population and $E_\textrm{min}, E_\textrm{max}$ correspond to the dynamically updated lowest and highest total energies in the population, respectively. 

With this definition, low (more negative) energy clusters have high fitness and high (less negative) energy clusters have low fitness. 

\subsubsection*{\bf Selection rule}
\label{sr}
 We use a ``roulette-wheel'' selection criterion \cite{roulette} with selection probability proportional to the value of the normalized fitness function. The idea is that the lower the total (or binding) energy (i.e., large negative value) of a certain configuration, the larger the probability to be chosen from the population. A cluster is picked at random and is selected for mating if its normalized fitness value ($\rho_i$) is greater than $\textrm{Rand}[0,1]$, a randomly generated number between 0 and 1 (i.e., if $\rho_i > \textrm{Rand}[0,1]$); where $\rho_i$ is the normalized fitness function defined in section-\ref{ff}. 
 
% A subtle problem is related to a possible poisoning of the selection pool with many structures that are all similar too each other. We have noticed that, frequently, a basin in the PES contains many local minima. These minima are different enough from each other to be judged as not {\em similar} by the geometric criterion defined below; on the other hand, some persistent topological feature is shared among all such minima. In such cases, the genetic pool may be flooded by a large number of alike structures, energetically close to the running GM, due to the high likelihood that mating among similar structures produces similar structures.
The above `` best-fit'' selection scheme can take significantly long time to reach another basin in the PES. In such situations, adding a little diversity by selecting one ``bad'' (high-energy) structure in the population is found to help in moving out to the next basin. Therefore, we define a complementary fitness function $\tilde{\rho}_j = (1-\rho_j)$ and we select one structure with high $\rho_i$ and another with high $\tilde{\rho}_j$. This choice, when the  the mixing ratio among different selection rules is optimized, greatly helps the convergence of the GA scheme and we also show how we optimized the mixing ratio among different selection rules.

\subsubsection*{\bf Crossover}
\label{sec:cross}
The crossover operator takes care of combining the two parent clusters selected as explained above. It is implemented as a modified version of the cut-and-splice crossover operator of Deaven and Ho.\cite{crossover} In our implementation of the cut-and-splice operation, first a random rotation is given (keeping the center of geometry of the cluster at the origin of the coordinate system) to both the parent clusters. Both clusters are then cut horizontally parallel to the $xy$-plane ($z=0$). Atoms with positive $z$-value are selected from one cluster and atoms with negative $z$-value are selected form the other cluster. These complementary fragments are spliced together. Importantly, this cut-and-splice operation does not ensure the preservation of the chosen cluster size (i.e., the total number of atoms) and the specific composition. We have adapted here three different kind of crossovers to maintain size and composition. 

(i) Crossover-1: Strictly speaking, this is a combined crossover and mutation (see below) step. After cut-and-splice we always maintain the same ordering of atoms that is given in the parent clusters. As an example, let us consider a small cluster like Mg$_2$O$_2$. In the parent cluster the ordering of atomic coordinate is given as Mg(1), O(2), Mg(3), O(4). When the cut-and-splice operation is applied, we get, for instance, a child with atomic coordinates from cluster one as Mg(1), Mg(3) (i.e., above the $xy$-plane) and that of from cluster two as Mg(3'), O(4') (i.e., below the $xy$-plane). Therefore, the entire atomic coordinates of the child are [Mg(1), Mg(3)], [Mg(3'), O(4')]. If this is the case, we replace the species of Mg(3) with O (without changing its coordinate) to impose the correct composition to the child. Thus the new ordering of atoms is: Mg(1), O(2), Mg(3), O(4) (i.e., the same composition as the parents). Therefore, it is possible that after the cut-and-splice operation a Mg atom of the parent cluster is replaced by an O atom in the child and vice versa.

% If the total number of atoms of the assembled child is not the one of the parents, we adapt the position of the cutting planes of both clusters by displacing them independently along the $z$ direction by small steps (the direction is defined by having shortage or rather overabundance of atoms in the child), until the assembly of the newly obtained child does give theright composition. 
% 
% After assemblage one or more pair of atoms may be found to be ``too close'' (1.5 \AA ~for the Mg-O system). This can happen only at the interface between the two pieces of the parents. In this case the two halves are pushed away along the $z$-axis until the minimum distance between pairs is satisfactory. Also this adjustment operation is regarded as {\em mutation} (see below).
%
% This approach is efficient in proposing new structures with the correct composition, but sometimes, due to the somewhat uncontrolled exchange of atomic species, it fails in maintaining certain winning features of the parent clusters. We have seen that it helps to proceed towards the GM quite fast, but near the GM it takes long time to find the actual one, especially for large clusters (number of atoms $\ge 40$). 
Also the total spin of the clusters is left free to evolve together with the spatial coordinates of the atoms. In this way we sample on equal footing the configurational space of atomic coordinates and the spin.
%Taking the cue from the method proposed by Gehrke and Reuter (2009)\cite{gehrke} 
The crossover of the spin coordinates is performed in the following way: when we create a new child by grabbing the atomic coordinates from the parents as explained above, we also make note of the atom-projected spin moments (via Hirshfeld partitioning of the electron density) for each atom. Such spin moments are given as initial moments of the individual atoms of the child. During the optimization process, these atom-projected moments are left free to change. 
%This choice, as opposed to the default assignation of the spin moments of the free atoms was found to greatly improve the convergence speed.

(ii) Crossover-2: In this procedure, after cut-and-splice we check whether the stoichiometry of the parents is maintained in the child. If it is maintained, we accept the child, otherwise we reject it and we iterate until until the child has the required stoichiometry.\cite{r15,r16} The advantage of this procedure is that it helps to maintain winning features of the parent molecule but most of the time it takes many iterations to obtain a valid child, even for a moderately sized cluster. In case one or more pairs of atom are too close, we adopt the same remedy as for crossover-1. The spin coordinates are taken care of the same as in the crossover-1 case.

(iii) Crossover-3: After re-orientation of the selected parent clusters we take all the metal (Mg) atoms from one parent molecule and all the oxygen atoms from another parent molecule. This crossover helps introducing diversity in the genetic pool, but  the rate of rejection during the assemblage of the child can be rather high due to the high likelihood that two atoms are too close.

\subsubsection*{\bf Mutation}
\label{mut}
After crossover, which generates a child, mutation is introduced.
Different mutation operators can be defined. We have adopted a) a translation between the two halves of the clusters (this is performed if atoms coming from the two different parents find themselves too close upon splicing of the two halves) and b) exchange of the atom species without perturbing their coordinates 

\subsubsection*{\bf Similarity of structures}
\label{ss}
In order to decide whether a newly found structure was already seen previously during the GA scan, after the local optimization we a) compare the energy of the new structure with that of all the others seen before and b) use a criterion based on the distances between all the atoms' pairs. In practice, we construct a coarse grained radial distribution function (rdf) of the clusters, consisting of 14 bins conveniently spaced. Each bin contains the (normalized) number of atom-pairs whose distance is between the distances that define the boundaries of the bin. For each cluster we have then a 14-dimensional rdf-array 
and the euclidean distance (i.e., the square root of the sum of the squared difference between corresponding elements in the two arrays) between the arrays arranged for two clusters is evaluated.
If this distance (note that it is a pure number) is greater than a convenient threshold (we used 0.01), then the structures are considered as different.

\subsubsection*{\bf Local optimization and early rejection scheme}
\label{cascade}

Although the geometry and the energy of the structures is not fully converged with PBE+vdW @ {\em low-level settings}, we have realized that there is a one-to-one mapping between the structures found at this level and those fully converged. In other words, if two structures are {\em similar} according to  PBE+vdW @ {\em low-level settings}, they are also at the PBE0+vdW @ {\em high-level settings} (see below). Furthermore, if a structure at the PBE0+vdW @ {\em high-level settings} is within $\sim 0.5$ eV from the running GM, with PBE+vdW @ {\em low-level settings} the structure is not further than 0.2 eV from the same running GM (with energy evaluated with PBE+vdW @ {\em low-level settings}). This implies that, with our conservative choice of not optimizing with the {\em high-level settings} structures that with {\em low-level settings} result positive to the {\em similarity} test or are more than 1.5 eV higher in energy than the current GM, we are not risking to reject structures that would eventually result in the GM or close to it. 

In the PBE+vdW, {\em high-level settings} optimization, atomic forces were converged to less than $10^{-6}$ eV/\AA. The grid settings where set to ``tight'' and the basis set was tier-2. In cascade, for the structure optimized in this way (i.e., without further optimization), we evaluated the PBE0+vdW energy with the tier-2 basis set. This energy is later used for the calculation of fitness of that particular cluster. 
The difference in binding energy between an isomer optimized with PBE0+vdW forces (tight / tier 1 / forces converged to 10$^{-5}$ eV/\AA) and the same optimized with PBE+vdW (tight / tier 2 / forces converged to 10$^{-5}$ eV/\AA), when the energy of both geometries is evaluated via PBE0+vdW (tight / tier 2), is small, i.e. at most 0.04 eV among all cases we checked. The computational cost of the PBE0+vdW further optimization would be thus not worthy (we estimated a gain of up to a factor 2 of overall computational time just by skipping the latter optimization)

\subsubsection*{\bf Parallelization}
The operation of selecting from the genetic pool two structures for the mating and the subsequent local optimization of the child, is an operation that can be performed at any moment also when a local optimization of a child is already running.
The algorithm is thus suitable for a very efficient parallelization. 

On top of FHI-aims parallelization (i.e. local optimization is run in parallel on an optimized number of CPUs) we add a second level of parallelization, i.e., we run at the same time several local optimizations, independently. The only communication among such replicas is the selection of the parents that is performed from a common genetic pool. The latter is also updated by each replica at the end of each local optimization.
The local optimizations run independently, i.e., each replica can start a new mating + local optimization cycle right after one is concluded; hence, there is no idling time between cycles.
Thus, we have $n$ local optimizations running in parallel, each requiring $p$ cores %\footnote{The number $p$ does not need to be the same for all replicas, but, since we deal with systems of the same size, this is the natural choice. Note that for FF-GA $p=1$}
, that fill the $n\times p$ cores required for the algorithm. The scaling behavior is about O($p^{1.5}$) with the number of cores for the local optimization part \cite{aimsp} . The number $p$ is indeed tuned in order to be sure that the speed-up is still O($p^{1.5}$) for the specific system. The scaling with respect to the $n$ replicas is linear, because the replicas are for the most of the time independent and only at the beginning and at the end of each local optimization, information is shared among the replicas.The first level of parallelization is performed within the FHI-aims code, by means of the MPI environment. The second level is script based: The total $n\times p$ number of cores is divided into $n$ groups, $n$ subdirectories are created and in each of them a cycle of local optimization job runs, each using $p$ cores.

\begin{table*}[b!]
\centering
{\bf \Large III. Performance of reaxFF\\}
\vspace{12pt}
\begin{tabular*}{\textwidth} {@{\extracolsep{\fill}}l | r | r | r | r | r |}
System & ReaxFF & PBE+vdW & PBE0+vdW & HSE06+vdW & Experiment \\
\hline
MgO bulk modulus [GPa] & 167 & 149~[1] & 169~[1] & 169~[1] & 165~[2] \\
MgO lattice constant [\AA] & 4.24 & 4.26~[1] & 4.21~[1] & 4.21~[1] & 4.21~[2] \\
O$_2$ binding energy [eV] & 5.38 & 6.23 & 5.37 & 5.31 & 5.22~[3] \\
O$_2$ bond distance [\AA] & 1.24 & 1.22 & 1.20 & 1.20 & 1.21~[4] \\
O$_2$ stretching freqency [cm$^{-1}$] & 1694 & 1555 & 1605 & 1663 & 1580~[4] \\
\end{tabular*}
\caption[]{Comparison of reaxFF with xc functionals on some of the properties used for training Mg and O parameters in reaxFF. The calculations for PBE+vdW, PBE0+vdW, and HSE06+vdW are done with tight-setting and tier-2 basis set. \\ \\
The reaxFF parameters for Mg and O were fit [5] to a quantum-mechanics training set including MgO$-$bulk and O$_2-$molecule properties. We checked that this parametrization yields for bulk MgO lattice constant within 1 \% and bulk modulus within 10 \% from the experimental values; similarly, for the MgO and O$_2$ dimer, it gives bond length and vibrational frequency within 1 \% and 10\% from the respective reference values. On this set of properties, reaxFF performs remarkably well, comparable to PBE+vdW, PBE0+vdW, and HSE06+vdW, but it fails very clearly for small clusters. In particular, ReaxFF yields a qualitatively good pre-scanning for stoichiometric clusters, while for non-stoichiometric ones it is far from desirable accuracy. This is not unexpected, as not even PBE is able to account for the charge redistributions associated with non-stoichiometry clusters. \\ \\
{\footnotesize $[1]$ J. Heyda and G. E. Scuseria, J. Chem. Phys., {\bf 121}, 1187 (2004). J. Paier, M. Marsman, K. Hummer, and G. Kresse, J. Chem. Phys. {\bf 124}, 154709 (2006). Erratum: J. Chem. Phys. {\bf 125}, 249901 (2006). \\
$[2]$ V. N. Staroverov, G. E. Scuseria, J. Tao, and J. P. Perdew, Phys. Rev. B {\bf 69}, 075102 (2004); Erratum: {\bf 78}, 239907(E) (2008). K. Marklund and S. A. Mahmoud, Physica Scripta. {\bf 3}, 75 (1971). \\
$[3]$ Chase Jr., M. W., NIST-JANAF Tables ``4th ed.'', J. Phys. Chem. Ref. Data Monogr. 9, Suppl. 1 (1998). Zero-point energy correction taken from Huber, K. P. and Herzberg, G., in {\em Molecular Spectra and Molecular Structure: Constants of Diatomic Molecules} (Van Nostrand Reinhold, New
York, 1979), Vol. 4. \\
$[4]$ K.P. Huber, G. Herzberg, Molecular Spectra and Molecular Structure. IV. Constants of Diatomic Molecules, Van Nostrand Reinhold Co. (1979).\\
$[5]$ R. Zhu, F. Janetzko, Y. Zhang, A. C. T. van Duin, W. A. Goddard and D. R. Salahub, Theor. Chem. Account, {\bf 120} 479 (2008).}
}
\label{T:comparison}
\end{table*}

\begin{figure*}[b!]
{\bf \Large IV. O$_2$-adsorption energy on MgO$_x$, with functionals corrected by the experimental value of O$_2$ binding energy\\}
\vspace{24pt}
\includegraphics[width=0.5\columnwidth,clip]{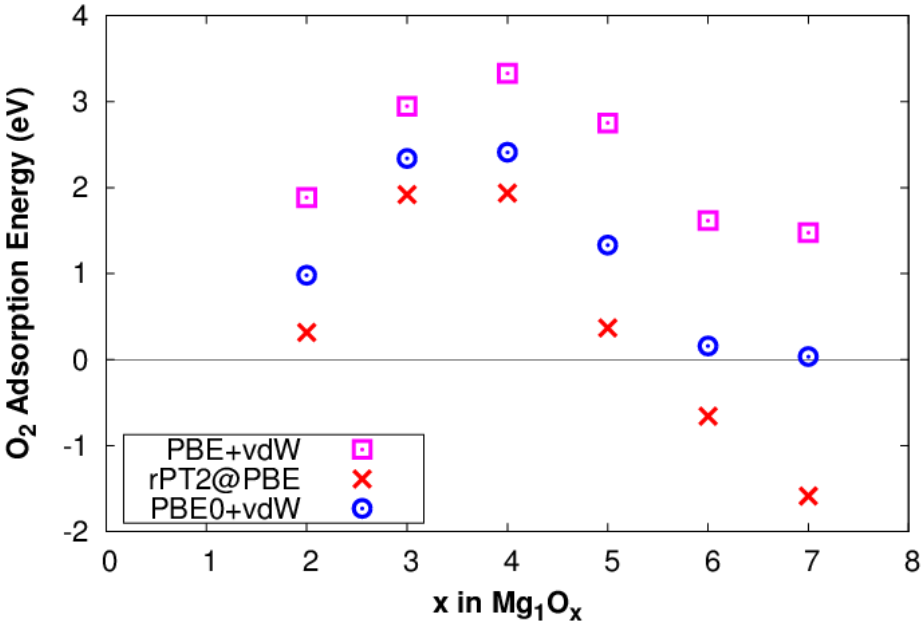}
\caption[]{O$_2$-adsorption energy of MgO$_x$ cluster using different xc functionals, calculated with O$_2$ molecule total energy corrected for the experimental value of O$_2$ binding energy (-5.21 eV [1]).\\
{\footnotesize$[1]$ Chase Jr., M. W., NIST-JANAF Tables ``4th ed.'', J. Phys. Chem. Ref. Data Monogr. 9, Suppl. 1 (1998). Zero-point energy correction taken from Huber, K. P. and Herzberg, G., in {\em Molecular Spectra and Molecular Structure: Constants of Diatomic Molecules} (Van Nostrand Reinhold, New York, 1979), Vol. 4.}}
\label{SM:be3}
\end{figure*}

\begin{figure*}[h!]
{\bf \Large V. O$_2$-adsorption energy on Mg$_2$O$_x$ and Mg$_3$O$_x$ clusters\\}
\vspace{24pt}
\includegraphics[width=0.45\columnwidth,clip]{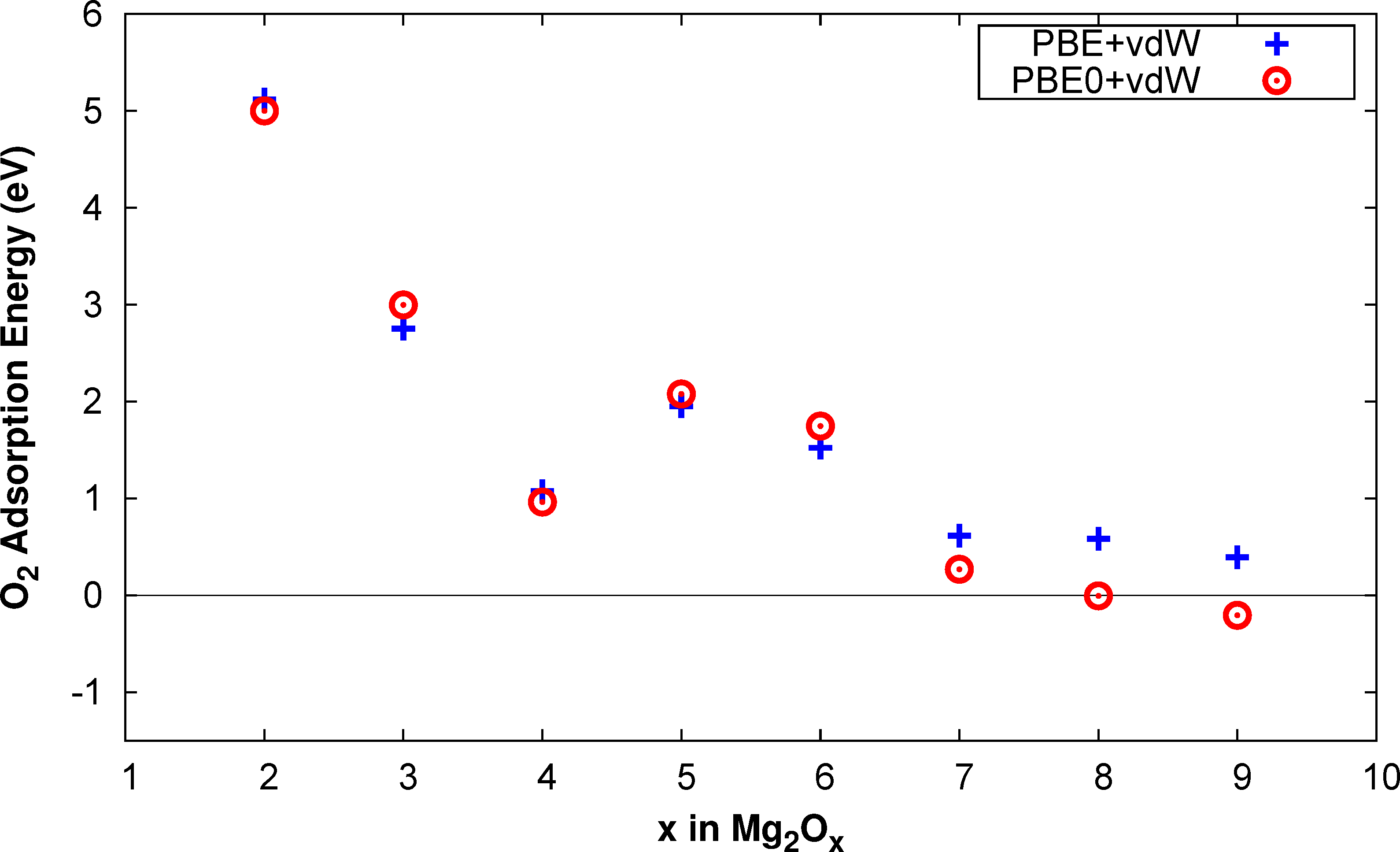}\\
\vspace{24pt}
\includegraphics[width=0.45\columnwidth,clip]{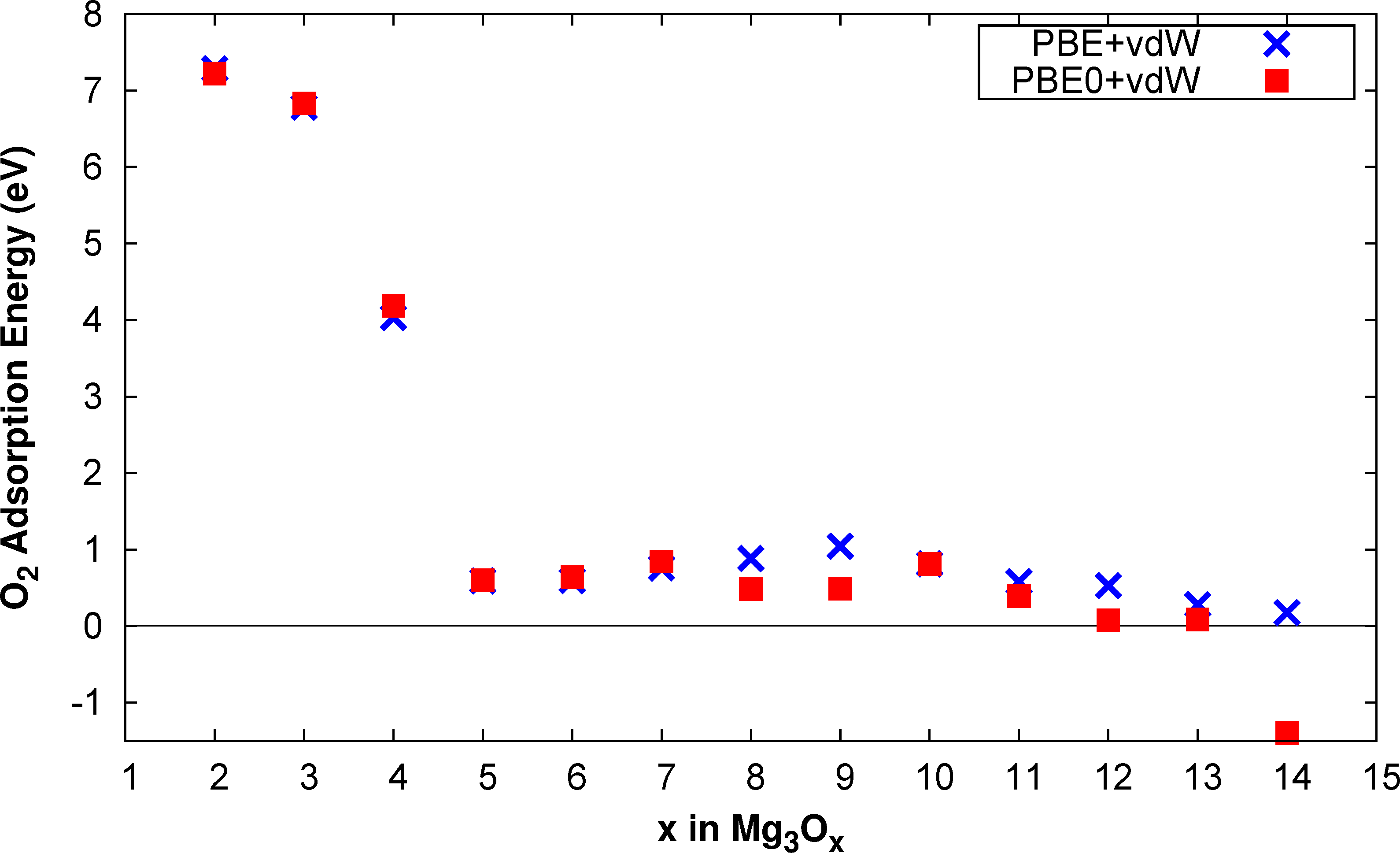}
\caption{O$_2$-adsorption energy of Mg$_2$O$_x$ (top) and Mg$_3$O$_x$ (bottom) cluster using PBE+vdW and PBE0+vdW functionals.}
\label{SM:adsorp4}
\end{figure*} 

\clearpage
\newpage

\begin{figure*}
\centering 
{\bf \Large VI. Mg$_2$O$_x$ phase diagrams with various functionals\\}
\vspace{24pt}
%\begin{table*}
\begin{tabular}{@{\extracolsep{\fill}}cc}
\def\subfigcapskip{12pt}
\subfigure[\, ReaxFF]{\includegraphics[width=0.4\textwidth,clip]{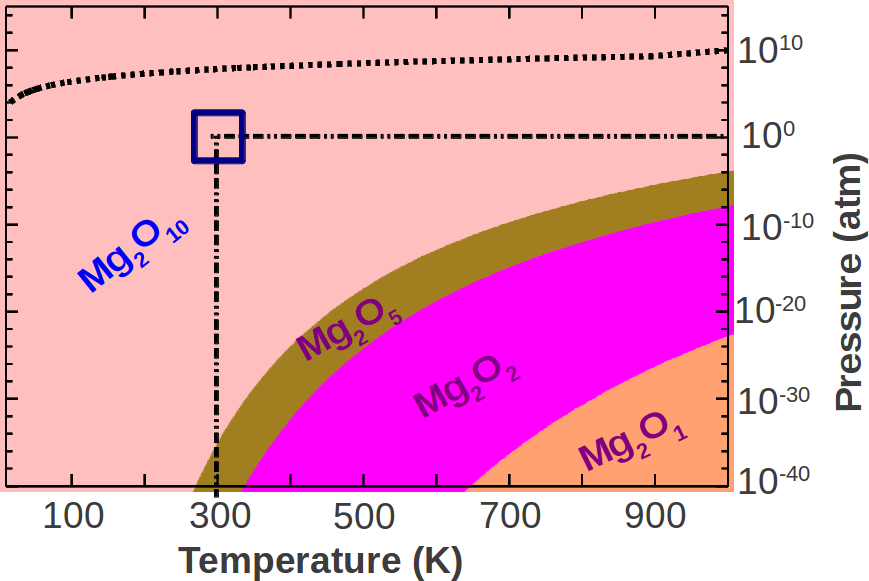}} &
\def\subfigcapskip{12pt}
\subfigure[\, PBE+vdW]{\includegraphics[width=0.4\textwidth,clip]{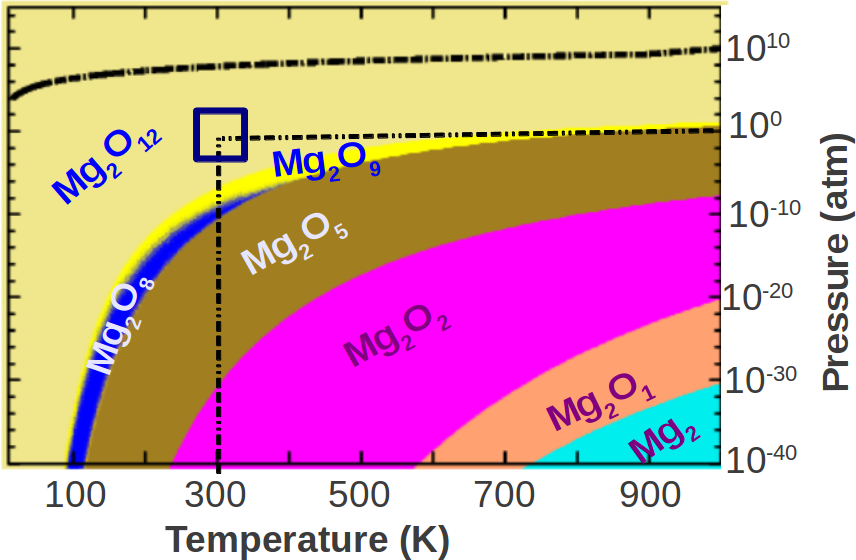}} \\
& \\
& \\
\def\subfigcapskip{12pt}
\subfigure[\, PBE0+vdW]{\includegraphics[width=0.4\textwidth,clip]{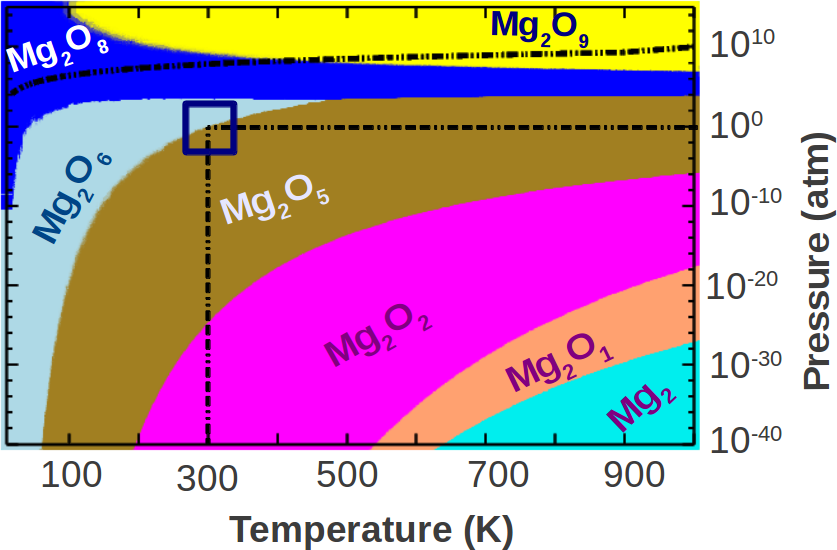}} &
\def\subfigcapskip{12pt}
\subfigure[\, PBE0 (no vdW)]{\includegraphics[width=0.4\textwidth,clip]{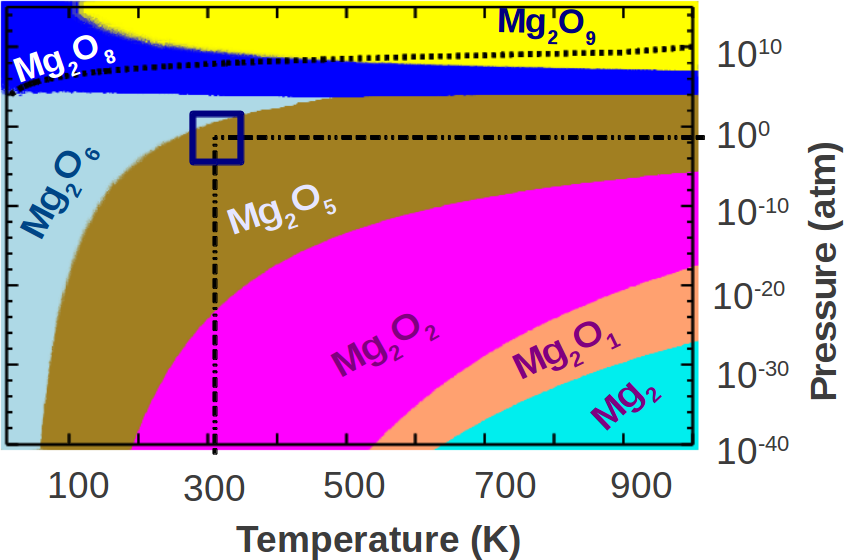}} \\
& \\
& \\
\def\subfigcapskip{12pt}
\subfigure[\, PBE0+vdW, exp. O$_2$ binding energy]{\includegraphics[width=0.4\textwidth,clip]{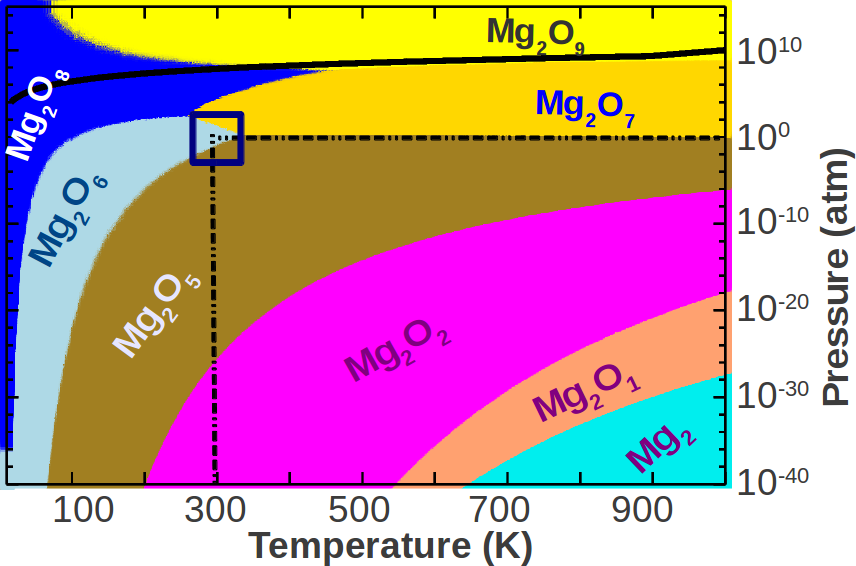}} &
\def\subfigcapskip{12pt}
\subfigure[\, rPT2@PBE]{\includegraphics[width=0.4\textwidth,clip]{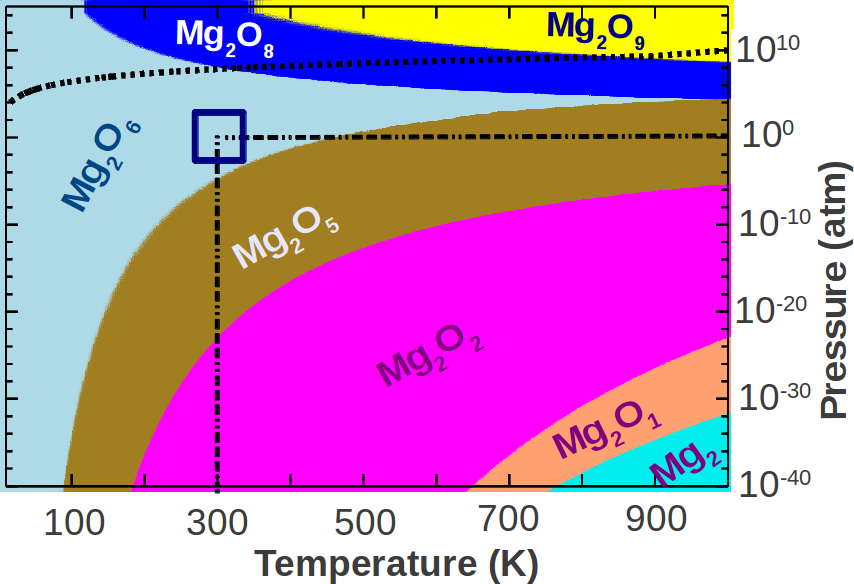}} \\
\end{tabular}
\caption{$(T, p_{\textrm{O}_2}) $ phase diagrams of Mg$_2$O$_x$ at different levels of theory. The square encompasses the region around normal conditions ($T$ = 300~K, $p_{\rm O_2}$ = 1~atm) and the dashed-dotted lines are guides for the eyes for identifying the point at normal conditions on the diagram.}
\end{figure*}

\begin{figure*}[t!]
\centering 
{\bf \Large VII. Effect of translational, rotational, vibrational contributions to the free energy on Mg$_2$O$_x$ phase diagram\\}
\vspace{24pt}
\begin{tabular}{@{\extracolsep{\fill}}cc}
\def\subfigcapskip{12pt}
\subfigure[\, Including $F^\textrm{translational}, F^\textrm{rotational}, F^\textrm{vibrational}$]{\includegraphics[width=0.39\textwidth,clip]{new-PT-Mg2Ox-PBE0.png}} &
\def\subfigcapskip{12pt}
\subfigure[\, Including only $F^\textrm{vibrational}$]{\includegraphics[width=0.4\textwidth,clip]{P-T-F-PBE0-only-vibs.png}} \\
& \\
& \\
\def\subfigcapskip{12pt}
\subfigure[\, Including $F^\textrm{translational}, F^\textrm{rotational}$]{\includegraphics[width=0.4\textwidth,clip]{P-T-F-PBE0-no-vibs.png}} &
\def\subfigcapskip{12pt}
\subfigure[\, Without $F^\textrm{translational}, F^\textrm{rotational}, F^\textrm{vibrational}$ ]{\includegraphics[width=0.4\textwidth,clip]{P-T-F-PBE0-Reuter.png}} \\
\end{tabular}
\caption{$(T, p_{\textrm{O}_2}) $ phase diagrams of Mg$_2$O$_x$ with different contributions in the free energy of the clusters.
The geometries are optimized with PBE+vdW and the electronic energy are calculated using PBE0+vdW. The sand-colored unlabeled regions are regions where different compositions (at least the adjacent ones) coexist (free energy of the coexisting species within 3 $k_\textrm{B}T$, see text).}
\end{figure*}

\begin{figure*}[b!]
\centering
{\bf \Large VIII. Effect of anharmonic contributions to configurational free energy\\}
\includegraphics[width=0.4\textwidth,clip]{fig-4.png}
%\includegraphics[width=0.4\textwidth,clip]{Mg4Ox-method1.png}
%\caption{Left: Anharmonic (AH) and harmonic (H) contribution to the Free energy at $p_{\textrm{O}_2} = 1$~atm of Mg$_4$O$_4$, Mg$_4$O$_{10}$, Mg$_4$O$_{12}$. Right: Phase diagram as in Fig. 3, with anharmonic configurational free energy.}
\caption{Anharmonic (AH) and harmonic (H) contribution to the Free energy at $p_{\textrm{O}_2} = 1$~atm of Mg$_4$O$_4$, Mg$_4$O$_{10}$, Mg$_4$O$_{12}$.}
\label{ah}
\end{figure*}

\begin{figure*}
\centering 
{\bf \Large IX. Examples of spin densities on non-stochiometric Mg$_M$O$_x$ \\}
\vspace{24pt}
\begin{tabular}{@{\extracolsep{\fill}}cc}
\def\subfigcapskip{12pt}
\subfigure[\,MgO$_4$, $\mathcal{M}=1$]{\includegraphics[width=0.2\textwidth,clip]{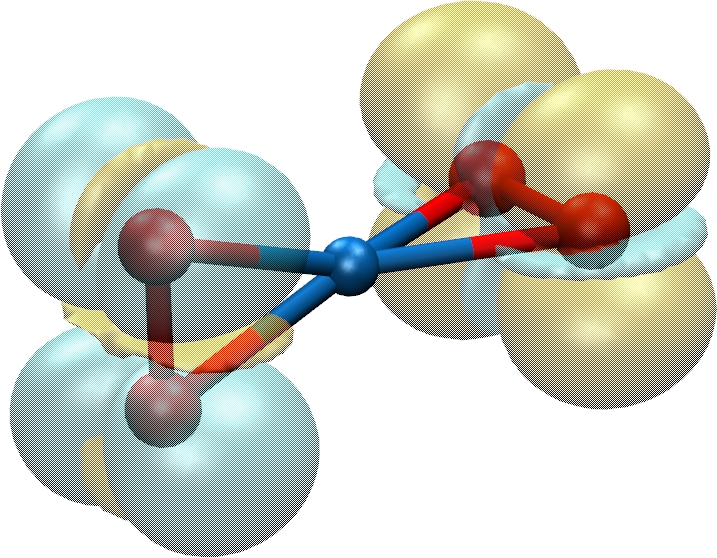}} &
\def\subfigcapskip{12pt}
\subfigure[\,MgO$_4$, $\mathcal{M}=3$]{\includegraphics[width=0.2\textwidth,clip]{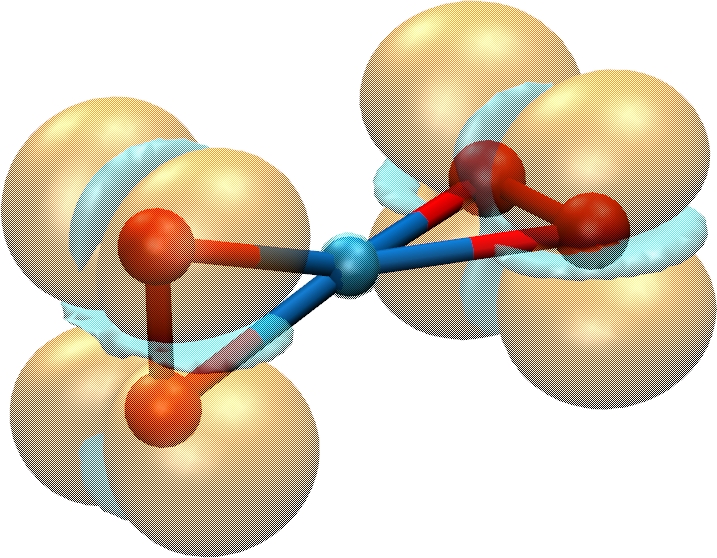}} \\
%& \\
%& \\
\def\subfigcapskip{12pt}
\subfigure[\,Mg$_3$O$_{10}$, $\mathcal{M}=3$]{\includegraphics[width=0.25\textwidth,clip]{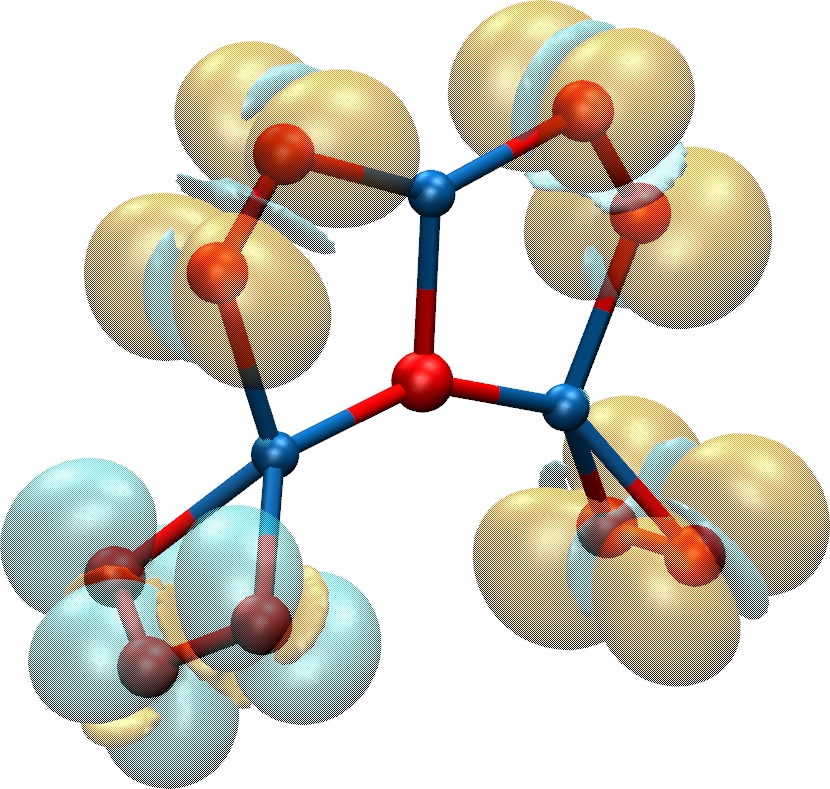}} &
\def\subfigcapskip{12pt}
\subfigure[\,Mg$_3$O$_{10}$, $\mathcal{M}=5$]{\includegraphics[width=0.25\textwidth,clip]{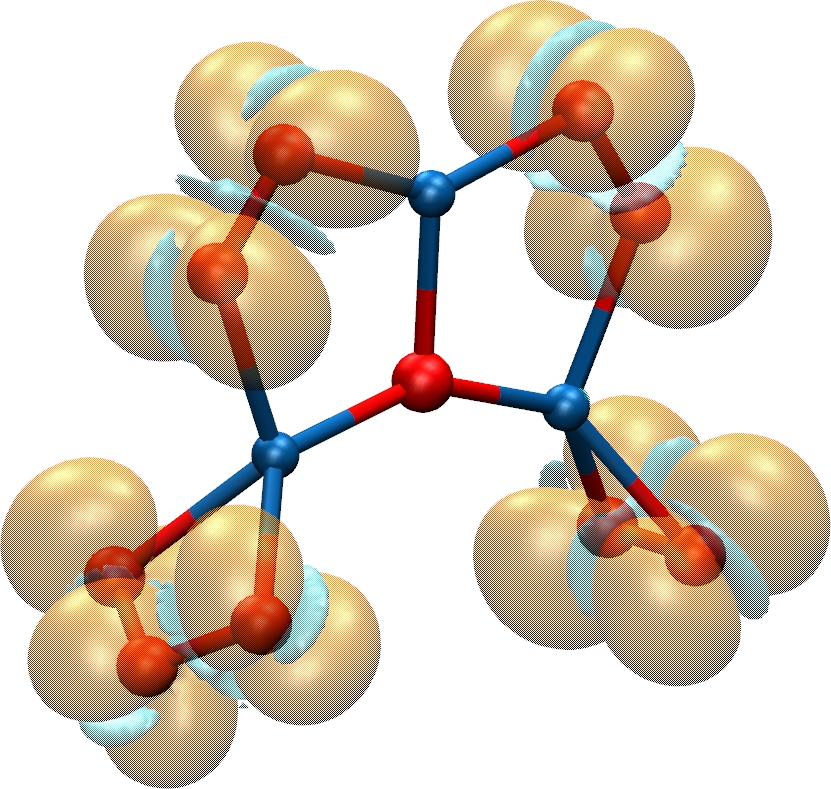}} \\
%& \\
%& \\
\def\subfigcapskip{12pt}
\subfigure[\,Mg$_4$O$_{12}$, $\mathcal{M}=3$]{\includegraphics[width=0.35\textwidth,clip]{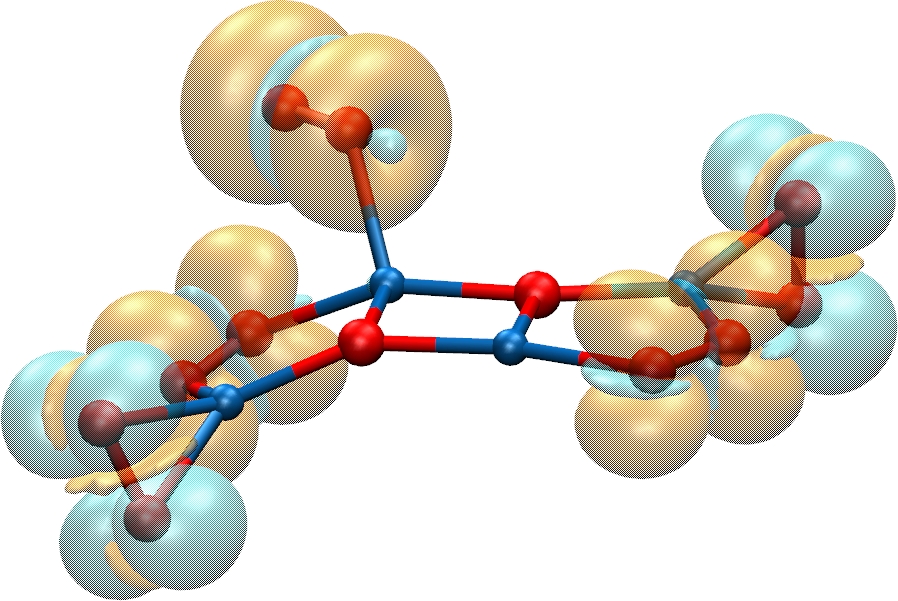}} &
\def\subfigcapskip{12pt}
\subfigure[\,Mg$_4$O$_{12}$, $\mathcal{M}=7$]{\includegraphics[width=0.35\textwidth,clip]{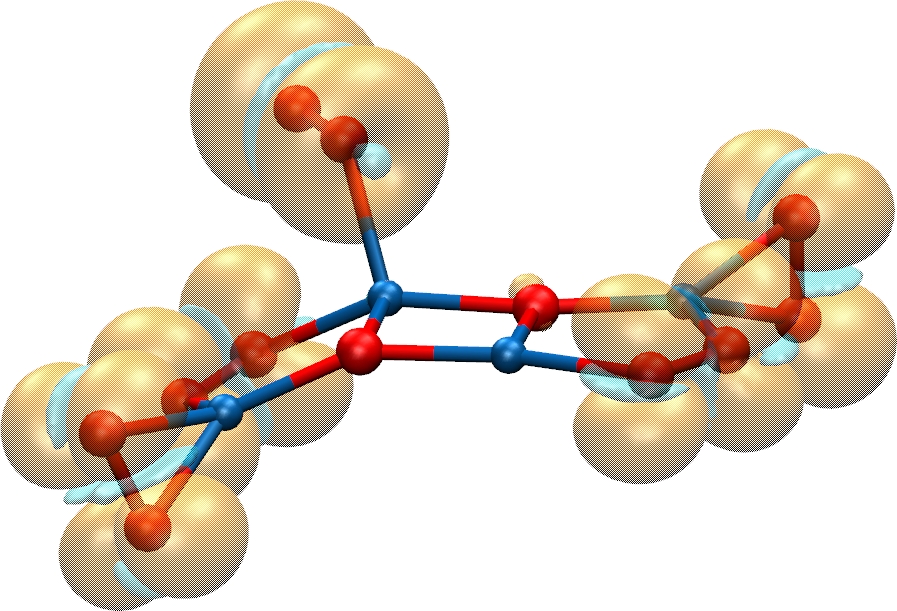}} \\
\end{tabular}
\caption{\footnotesize{Spin density (difference between the majority-spin and the minority-spin electronic densities) for three representative clusters: MgO$_4$ ($\mathcal{M}=1$ and $\mathcal{M}=3$), Mg$_3$O$_{10}$ ($\mathcal{M}=3,5$), and Mg$_4$O$_{12}$ ($\mathcal{M}=3,7$). Red (blue) spheres represent O (Mg) nuclei. Orange (blue) lobes are spin-density isosurfaces at 0.01 (-0.01) $e^-/$\AA$^3$ level. The two spin states of MgO$_4$ ($\mathcal{M}=1$ and $\mathcal{M}=3$) are nearly degenerate in energy (at the PBE0+vdW level). Mg$_3$O$_{10}$ ($\mathcal{M}=3$, isoenergetic with $\mathcal{M}=1$) contains an O$_3$ moiety, which hosts one unpaired electron in the same fashion as the O$_2$ moieties. Having Mg$_3$O$_{10}$ four O$_2$/O$_3$ moieties, its largest low-energy (0.02 eV higher than the singlet ground state) multiplicity is $\mathcal{M}=5$. With its five O$_2$ moieties, Mg$_4$O$_{12}$ shows at 0.05 eV from the singlet ground state an isomer with spin multiplicity $\mathcal{M}=7$. The shown $\mathcal{M}=3$ state is 0.03 eV from the singlet ground state. The off-plane O$_2$ moiety hosts two unpaired electrons.}}
\label{Fig:spindens}
\end{figure*}